# Multi-Criteria Assessment of Shape Quality in CAD Systems of the Future

Valerijan Muftejev[1,2][0000-0003-4352-3381], Rushan Ziatdinov[3,*][0000-0002-3822-4275], and Rifkat Nabiyev[4][0000-0002-0920-6780]

[1] Department of the Fundamentals of Mechanisms and Machines Design, Ufa State Aviation Technical University, Ufa, Russian Federation
muftejev@mail.ru
http://www.spliner.ru

[2] C3D Labs, Altufevskoe Shosse 1, Office 112, 127106 Moscow, Russian Federation

[3] Department of Industrial Engineering, Keimyung University, Daegu, South Korea
ziatdinov@kmu.ac.kr, ziatdinov.rushan@gmail.com
http://www.ziatdinov-lab.com

[4] Department of Management and Service in Technical Systems, Ufa State Petroleum Technological University, Ufa, Russian Federation
dizain55@yandex.ru

**Abstract.** Unlike many other works, where authors are usually focused on one or two quality criteria, the current manuscript, which is a generalization of the article [35] published in Russian, offers a multi-criteria approach to the assessment of the shape quality of curves that constitute component parts of the surfaces used for the computer modelling of object shapes in various types of design. Based on the analysis of point particle motion along a curved path, requirements for the quality of functional curves are proposed: a high order of smoothness, a minimum number of curvature extrema, minimization of the maximum value of curvature and its variation rate, minimization of the potential energy of the curve, and aesthetic analysis from the standpoint of the laws of technical aesthetics. The authors do not set themselves the task of giving a simple and precise mathematical definition of such curves. On the contrary, this category can include various curves that meet certain quality criteria, the refinement and addition of which is possible in the near future. Engineering practice shows that quality criteria can change over time, which does not diminish the need to develop multi-criteria methods for assessing the quality of geometric shapes. Technical issues faced during edge rounding in 3D models that affect the quality of industrial design product shape have been reviewed as an example of the imperfection of existing CAD systems.

'*There is no such thing as an unsolvable problem.*'
Sergei Korolev

## 1 Introduction

In engineering design, plane and spatial curves that specify certain functional characteristics of an object are known as *functional curves* [1]. Among the functional curves, a subclass of *engineering curves* may be distinguished; such curves prescribe some design characteristic of an object in a single optimum way. These kinds of curves, for instance, include the Archimedean spiral used for shaping the profile of gear teeth, as well as the *brachistochrone* – the fastest descent curve for transporting items [2]. A

---

*catenary* used for dome and hanging structure surface design, as well as the *clothoid*, which is used to design smooth roadway transitions [3], can also serve as examples of engineering curves.

Engineering curves are widely used for solving various tasks and issues faced in different branches of technology and industry. Below are some examples:

1. The wing profile of an aeroplane creates lift; therefore, when designing a profile curve it is necessary to maximize the lift while minimizing the drag.
2. A road ensures comfortable and safe vehicle driving at a given speed, which is why maximum road smoothness should be achieved within given limits and restrictions.
3. A cam profile defines the movement of a pusher with a valve to ensure a necessary gas distribution pattern; therefore, its design should ensure smooth, impactless valve movement.
4. The external surface of a vehicle body and the curved architectural shapes of a building can also be functional surfaces if one views aesthetics and beauty as a design property of a product that determines its usability.

Plane free-form functional curves can be locally convex (with a curvature function of constant sign) and may feature points of inflection (areas with a curvature function of variable sign). Furthermore, functional curves may be spatial and may, therefore, have torsion. Those interested in plane curves are advised to read a well-known reference book by Savelov [4].

In their previous works, the authors have defined basic and supplemental quality requirements for functional curves using smoothness criteria applicable to technical objects [1], [5-7]. Several studies have been dedicated to the development of modelling methods for aesthetic curves and their quality assessment from the standpoint of the laws of technical aesthetics [8-9]. In the present manuscript, these results have been clarified, expanded and systematized. Functional and aesthetic curves have been viewed from a unified standpoint; common quality assessment criteria have been suggested. Methods of modelling the curves meeting these requirements have been reviewed. Key points of the methods crucial to authors' priorities have been described in detail in the manuscript.

## 2    Quality of Geometric Shapes

Regardless of the specifics of the items designed, one can derive universal requirements for the quality of geometric shapes arising from free-form functional curves. This section offers a general list of the quality requirements for functional curve shapes that are invariant as regards the specifics of an item design. Additionally, readers who are interested in high-quality shapes are recommended to try the FairCurveModeler app, which can be accessed online at http://fair-nurbs.ru/FairCurveModeler3D.aspx.

### 2.1    Order of Smoothness Not Less Than 4

Smoothness is a property of a function or a geometric figure (a curve, a surface, etc.) indicating that this function can be differentiated or that each point of the given figure has surroundings that can be defined using differentiable functions. Different types of design use splines of different orders of smoothness. For example, when designing road routes, clothoid splines are used and smoothness of at least the 2nd order is ensured. For profiling the camshaft cam of high-speed engines, smoothness of at least the 3rd order is required; therefore, the profile design begins with drawing a smooth graph of the 3rd order derivative [10]. To ensure continuity of the torsion function when modelling spatial curves, the curve must have 3rd order smoothness. A spatial curve with smooth torsion should have 4th order smoothness, which follows from the analysis of the spatial curvilinear trajectory of the point particle [11].

By analogy with the concept of jerk (a quick, sharp, sudden movement) for plane curves, meaning a sharp change in the rate of change of curvature, we can introduce the concept of jerk for a sharp change in the rate of change of torsion. Only spline curves of the 5th degree or higher (at least 4th order of smoothness) provide a smooth change in torsion and can be used to model functional curves. A surface can be drawn from a network of plane curves. However, the jets of a medium (air along the wing, water along the propeller blade, soil along the plough blade) do not flow around an object, in the general case, along planar curves; they flow around its surface along spatial trajectories with torsion. If the jet trajectory does not have smoothness 3, then the discontinuities of the derivatives of the 3rd order will inevitably cause discontinuities in the torque function. Pulsating moments of forces (sharp pulsations at smoothness order 2 and smoother pulsations at order 3) acting on spatial jets of the medium cause flow pulsation, which, inevitably, increases the dynamic resistance of the surface to the movement of the medium flow. Therefore, planar or spatial curves in the curve network must also have a smoothness order of 4 or higher. In addition, the formula for defining a surface on a network of curves must provide an order of 4 or higher for any isoparametric curve of a surface.

## 2.2 Absence or Minimum Quantity of Curvature Extrema

The smoothness of the line also depends on the shape of the graph of the change in curvature along the length of the motion line. According to the basic dynamics equation [11], oscillations of the curvature function will cause the pulsation of centrifugal forces acting on the point particle. Therefore, the section of the motion line must have a minimum number of extrema of the curvature or a minimum number of vertices of the curve. For instance, the presence of redundant extrema of the curvature in the shape of the designed item may result in the following deviations:

1. It can cause undue runout of the pusher that ultimately leads to premature mechanism wear.
2. It can cause soil build-up on a plough section with curvature concentration at the soil movement trajectory, which leads to increased resistance of the plough and ultimately increases the energy intensity of the ploughing process [12].
3. Their presence on the aerodynamic profile can lead to excessive pulsation of the medium flowing around the profile, which increases the drag on the profile and can cause a flow stall, as well as an increase in the pressure force on the profile.
4. It can cause the need for excessive braking and acceleration, which would ultimately increase the energy required for the movement of a vehicle [13].
5. Their presence on the curves of vehicle body part surfaces and architectural forms can result in distorting mirror effects [14].
6. They may cause incorrect visual perception of computer graphics and CAD objects [15].

## 2.3 Small Values of Curvature Variation and Its Variation Rate

In some applications, a requirement is introduced to minimize the variation in the curvature. For example, such limitation to the minimum value of the curvature radius (max curvature) is introduced naturally during a road design, where the minimum bend radius is limited based on the allowed vehicle speed [16-17].

An important quality attribute of a curve is the rate of variation in its curvature. When designing a road route, this attribute defines the rate of centrifugal force increase impacting a vehicle at bends in the road, and it is easily controlled through applying the segments of the clothoid with a linear curvature function variation [16-17].

## 2.4 Small Value of the Potential Energy of the Curve

The curve with a minimum value of potential energy is called an *elastica* [18]:

$$E_{MEC} = \int_{l_0}^{l_1} \kappa^2(s)\, ds \to min \qquad (1)$$

It is an axis line of a deformed elastic bar between two fixed endpoints. The quality of elasticas has been proven by the centuries-old shipbuilding experience. Elastic bars (physical splines) have been used in the profile lofting of transverse frame ribs, buttocks and water lines in the design and construction of marine vessels and, later on, in the production of automobiles and aircraft.

Mathematically accurate modelling of the contour of a curved physical spline is used in the KURGLA curve modelling program for the AUTOKON ship design system [14], [19]. In one of the KURGLA algorithms, a virtual physical spline is approximated by clothoid segments. According to [20], the curvature between the fixed points of a physical spline varies linearly as is the case with a clothoid.

The curve smoothness is believed to be directly related to the potential energy of such a curve. The need to choose a functional curve with a small potential energy value is justified by the following assumption. When an object with a functional surface moves at a high speed, the medium flowing around the object behaves like an elastic body, and less pressure will be required to deform the elastic medium along streamlines with less potential energy. When a point particle moves along a concave curved path, with friction taken into account the work spent on moving it will be less with a lower value of the potential energy of the moving path [1]. This situation is also true for the point particle movement along a curvilinear plane trajectory given the friction.

The development of scientific visualization methods opens up new possibilities for the mathematical modelling of geometric shapes[1]. There is an opportunity to study polynomial and nonlinear splines by means of a computational experiment and, as a result, obtain high-quality visualizations with high resolution. In such visualizations, points are determined by pixels, and calculating an area with a resolution of 100 × 100 pixels can take several minutes. In [21], visualizations were obtained for the potential energy function of a quadratic Bézier curve with a monotonic curvature function.

### 2.5 Aesthetic Analysis from the Standpoint of the Laws of Technical Aesthetics, Based on Eleven Criteria

**What is a Beautiful Shape?**

The beauty of the shape of an industrial product is a *measure of quality perfection* expressed in visually perceived characteristics (geometry, colour, style, etc.) of the shape, resulting from objectively acting conditions: function, design, properties of materials, compliance with human factors (anthropometric, ergonomic, aesthetic, etc.) and formed by means of design shaping (proportioning, compositional balance, tectonic pattern, volumetric spatial organization, colour harmony, etc.).

The expressed beauty of the shape of an industrial product effectively embodies the content (function, purpose) and causes a positive emotional and psychological reaction in a person.

**Necessity of Aesthetic Analysis**

Design practice carried out in the field of high-tech industrial production through the mathematical modelling of industrial products and the evaluation at the production site of manufactured industrial samples is needed to ensure maximum efficiency, economy and performance of the product throughout its entire life cycle. However, the performance of an industrial sample is not limited to technical characteristics only. The product functions in all the variety of its relations with a person who reacts to objective stimuli and evaluates them not only on the basis of rational judgments and conclusions, but also in terms of the emotional–sensual attitude to the world. In that sense, a future design solution should include not only a rational but also an aesthetical feasibility model at the pre-design analysis stage already. Such a dialectical unity is transformed into a harmoniously integrated image that gives rise to a motive: an incentive for the emotional perception of the formal qualities of a shape to be a factor in the desire to reveal the useful qualities of the product, which in general will determine the value judgment about it.

From the standpoint of technical aesthetics, the achievement of the unity of rational and emotional aspects in the image of a product is determined by the objective laws of shaping. In terms of their objectification, it is important to identify the characteristics of the primary elements of a shape, the content of which in many respects sets the qualitative properties of a product design solution.

---

[1]  'Visualization Methods in the Mathematical Modeling of Interpolating Curves with Monotonic Curvature Function.' YouTube, uploaded by Geometric Analysis, December 22, 2016, https://www.youtube.com/watch?v=xlUFKVageu8

The quality assessment of a curve, including from the standpoint of the laws of technical aesthetics, must be carried out according to the proposed objective method for assessing smoothness. A designer who is not restricted by the need to search for an engineering curve can model free-form curves using given Hermite data.

**Aesthetic Analysis of Quadratic Bézier Curves**

A planar Bézier curve[2] was used as such a primary element in [8-9]. Its geometric properties were analysed and evaluated from the standpoint of their aesthetic feasibility together with the ability to meet the efficiency requirements. The principle of 'structural unity of a shape' was used as a basis for scrutiny of the formative, plastic and expressive properties of the curves. The shape-forming features of the geometry of the curves were evaluated according to the following eleven criteria: *conciseness-integrity, expressiveness, proportional consistency, compositional balance, structural organization, imagery, efficiency, dynamism, scale, plasticity and harmony* (more detailed information on these criteria can be found in [38-39]).

Art–design analysis of available samples of Bézier curves revealed regularities of shaping at the level of geometric features of the curve based on the fundamental principles of the volumetric and spatial organization of the shape [9]. It is also worthwhile to note that the data objectification was performed by employing a questionnaire. Its goal was to differentiate curve structure assessment by professionals creating product samples and their design, as well as the emotional and sensual response of ordinary consumers to the features of the Bézier curves offered to them for appraisal. Art and design analysis does not exhaust all aspects of the issue under consideration and offers the prospect of further development.

The study in [9] contains a detailed aesthetic analysis of 24 segments of Bézier curves of the second order, 8 of which had a monotonic curvature function. Each of the above eleven criteria was assessed according to a seven-point scale from -3 to 3 (maximum degree, medium degree, minimum degree, no criterion, minimum deviation, medium deviation, maximum deviation). The analysis has shown that in four segments of the Bézier curves with a monotonic curvature function the *rounded average value of fairness* (RAVF) for all criteria is 0; in three segments it is 1, i.e. the criteria are of minimum degree, and in one curve segment the criteria breach has been identified.

To support the authors' conclusions [9], a questionnaire was administered in one of the leading schools in Istanbul, Turkey to 240 teenagers from 14 to 17 years of age to investigate the 'aesthetic feasibility' of different segments of Bézier curves, and its results completely matched those of the authors' ones. The choice of the age group was based on consideration of teenagers' psychological peculiarities in forming an emotional picture of the world, with the characteristic absence of psychological dependence of children's consciousness on professional dogmas, norms and instructions inherent in the adult audience. In this sense, this age group makes it possible to give answers to the questionnaires based more on intuition and sensual perception than on rational judgments and inferences, which is necessary to objectivize the results of the questionnaire.

Bézier curves having a monotonic curvature function (class A Bézier curves) are often considered *aesthetic (fair) curves* [43], although their aesthetic analysis has never been performed. A detailed aesthetic analysis carried out in [9] showed that this statement is erroneous.

The authors of the current work believe that assessment using the criteria of smoothness is a priority. An expert assessment from the standpoint of the laws of technical aesthetics is valid only after an assessment of smoothness or in the absence of the possibility of such an analysis.

**Natural Beauty of Spiral Curves**

There is another approach to assessing the aesthetics of a curve that is based on the mathematical characteristics of shapes found in real-world objects (e.g. the outlines of butterfly wings) [22-23]. To generate beautiful (aesthetic) shapes, the so-called *log-aesthetic curves*[3] – which have a linear graph of curvature in a logarithmic scale – are suggested [24-26]. Many well-known spirals [42], including a clothoid, are special cases of this class of curves. The most generalized class of curves with a monotonic curvature function, called *superspirals*, was introduced in [27] and studied via similarity geometry in recent works [40-41]. Equations of these curves are expressed through Gaussian hypergeometric functions and are numerically integrated by adaptive integration methods such as the Gauss–Kronrod method.

---

[2]  'Bernstein Polynomials and Bernstein-Bézier Curves.' YouTube, uploaded by Rushan Ziatdinov, June 30, 2015, https://www.youtube.com/watch?v=AL0vcsLlYp4

## 3 Class F Curves

### 3.1 Modelling Methods

Thus, building a very smooth trajectory of the motion requires a minimum number of reference points of the generated spline motion trajectory and a high level of smoothness of at least the 4th order, smooth torsion of the spatial curve, restriction of the maximum value of curvature and the variation rate of curvature, and minimization of the potential energy function. Functional curves satisfying these requirements are called *class F curves*[4,5] [29], [36]. The authors do not set themselves the task of giving a simple and precise mathematical definition of such curves. On the contrary, this category can include various curves that meet certain quality criteria, the refinement and addition of which is possible in the near future. Engineering practice shows that quality criteria can change over time, which does not diminish the need to develop multi-criteria methods for assessing the quality of geometric shapes.

In the Russian language, a curve model is called the *determinant* [29], which consists of the geometric part plus the algorithm for generating curve points or the procedure for constructing an approximating spline. The geometric part of the determinant can be considered the geometric determinant of the curve. The most common and natural forms of the geometric determinant are the sets of points (the type of the polyline vertices) or the set of tangent lines (namely, the form of the tangent polyline). Also, the so-called control spline polygons of NURBS curves are used in the applied geometry. Different types of geometric determinants have their own advantages and disadvantages. The incidence line enables accurate positioning of the curve, the tangent line uniquely and accurately sets the shape of the modelled curve, and the NURBS S-polygon of the high-degree curve enables local change of the shape of the curve, guaranteeing high-quality spatial curves according to the criteria for smooth curvature and torsion.

A general curve modelling algorithm includes the following steps [5]:

1. Sketching the curve. Preliminary information about a curve may be specified as a) a curve gauge and its digital representation, b) multiple points captured from a full-scale replica using a measurement device, c) a line drawn by an engineer on paper or on a screen and recorded as a digital set of points, d) a digital set of points the curve should cross, or e) a fixed analytical curve.

2. Plotting a geometric determinant of a curve defining the geometric structure of the curve on the sketch.

3. Isogeometric approximation of the geometric determinant through developing an analytical (or piecewise-analytical) curve of a given class or plotting the results of an algorithm for generating curve points based on the given parameters of the geometric determinant.

4. Transition to another type of curve determinant by equivalent transformation or by isogeometric approximation of the curve determinant, editing it using the parameters of the new geometric

---

[3] 'Interactive Aesthetic Curve Segments.' Personal webpage of Norimasa Yoshida, 2006, http://www.yoshida-lab.net/aesthetic/pg2006iacs.wmv
[4] Authors should not confuse this term with so-called *F-curves* proposed by Ferguson [37].
[5]  'Methods of high-quality surface modeling.' YouTube, uploaded by Rushan Ziatdinov, August 22, 2018, https://www.youtube.com/watch?v=YuNTTlz7K70 [in Russian].

   determinant.
5. Transition to another type of curve determinant by equivalent transformation or by isogeometric approximation of the curve determinant to solve metric and positional tasks in CAD systems. In this case, the new curve determinant is called a curve pattern.

### 3.2   Absence or Minimum Number of Curvature Extrema

General requirements for curve modelling methods are formulated in [6], [14] and [29-30]. These requirements include dimensional stability or isogeometry, invariance under affine and projective transformations, high quality according to the criteria of smoothness and aesthetics, flexibility, instrumental diversity and a possibility of using analytical curves.

## 4   Comparative Assessment of Curves

An unbiased comparison of curves requires that they be based on the same Hermite data and subjected to comparative analysis against the smoothness criteria. When comparing two curves constructed using the same Hermite data, the number of curvature extrema is checked, and the curve with the larger number is rejected. Then the order of smoothness is compared, and the curve with less smoothness is rejected. Further on, the curves are compared using the value of potential energy. The last stage of this assessment can be an aesthetic analysis from the standpoint of the laws of technical aesthetics.

## 5   Analysis of the Functional Capabilities of CAD Systems in the Context of Quality Assessment for the Surfaces of Industrial Design Products

Today, computer-aided design systems available on the market provide the industrial designer with tools to create digital prototypes of products at the required level of accuracy. Principally, they are generated through solid modelling. However, it is known that this method does not to fully solve the issue of creating the complex geometry of a form. Therefore, for this purpose, surface modelling is used. In this regard, many CAD systems feature special modules focused on creating products with complex surface geometry. However, for the productive use of this toolkit, an industrial designer needs to have a sufficiently deep understanding of the theoretical aspects and patterns of the technology used and needs to spend much time searching for solutions to purely technical problems while using a software product. Otherwise, the existing product design concept cannot be fully developed by the creator in the software environment, which negatively affects the entire design process. A particularly urgent problem is the rounding of the edges of the 3D model in the process of modelling a design product. In many CAD systems, the issue of automatic and high-quality edge rounding has not yet been resolved. But its solution would save designers from the CAD mathematical apparatus and allow them to focus on the process of finding the optimal form of a product designed. That is, to proceed to the tasks of their immediate area of expertise. Let us look at a few examples that visually illustrate the issue of rounding edges in CAD systems such as Rhinoceros [6], Altair Inspire Studio [7], ANSYS SpaceClaim [8] and Autodesk Inventor Professional [9]. These programs were given the task of rounding, in automatic mode, all the edges of a solid body consisting of two mutually perpendicular parallelepipeds making contact on their faces (Fig. 1).

---

[6] https://www.rhino3d.com/
[7] https://solidthinking.com/product/inspire-studio/
[8] http://www.spaceclaim.com/en/default.aspx
[9] https://www.autodesk.com/products/inventor/overview

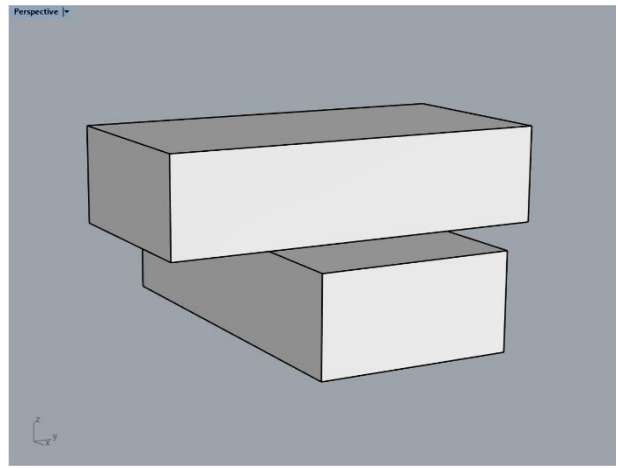

**Fig. 1.** A solid body used for edge rounding.

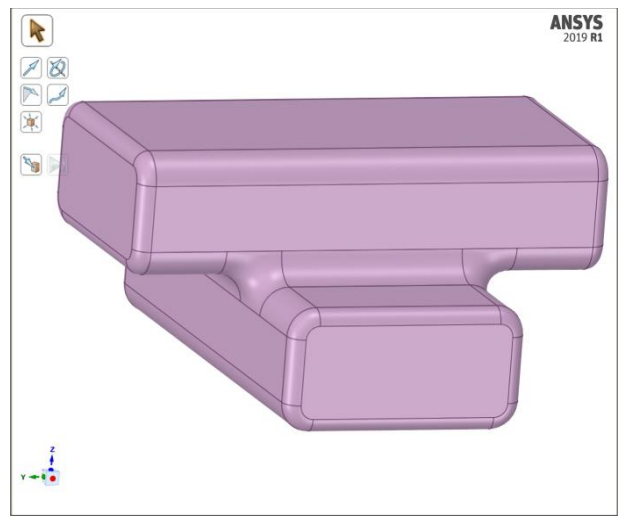

**Fig. 2.** Automatic edge rounding result in ANSYS SpaceClaim.

The edges have been rounded with the continuity of curvature $G^2$. According to [33], '$G^2$ is known to connect profiled curved surfaces with the curvature continuity to the boundary surfaces. With this connection type, one curve transfers to the other and the end point of the former coincides with the starting point of the latter one. Besides, tangent angles and radii at these points coincide.' The programs under consideration formally coped with the task of forming secondary surfaces, mating with primary surfaces with continuity $G^2$ (Figs. 2, 3, 4), except for Rhinoceros. This application failed the task of rounding edges in the area of their intersection (Fig. 5). A peculiarity of the edge rounding in Autodesk Inventor Professional 2020 was the creation of a set of extra surfaces, making the topology of the object more complex (Fig. 4).

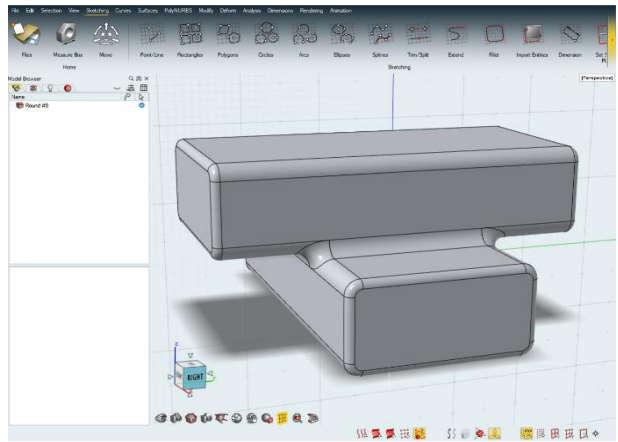

**Fig. 3.** Automatic edge rounding result in Inspire Studio.

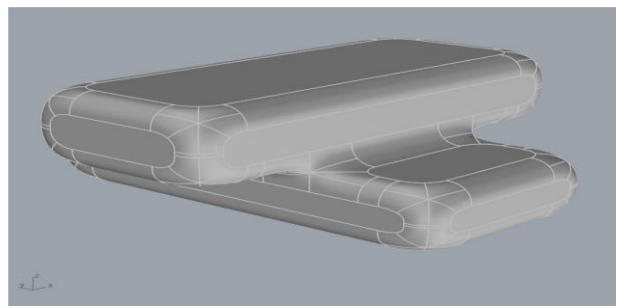

**Fig. 4.** Automatic edge rounding result in Autodesk Inventor Professional 2020.

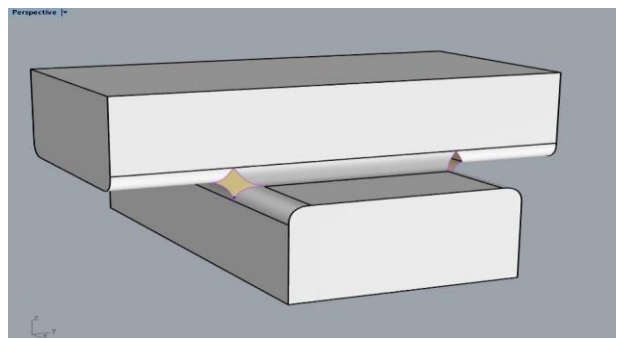

**Fig. 5.** Automatic edge rounding result in Rhinoceros 6.

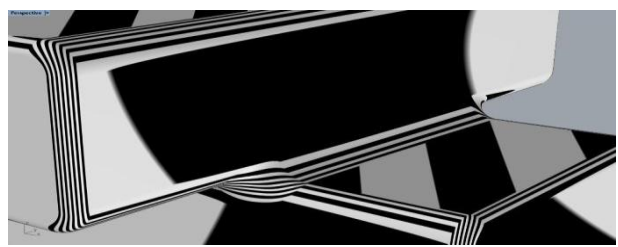

**Fig. 6.** Analysis of the surface of the designed object using zebra lines in Inspire Studio.

It is known that an industrial design product should embody useful beauty in its form, the characteristics of which express the functional expediency of the product. From a technical aspect, one of the conditions for creating such a product is the visual purity of its shape, expressed in the uniform movement of light flare over its surface. This is especially true for products in which class A and F surfaces are used.

In a software environment, the behaviour of light flare can be predicted by surface analysis using several methods. Here zebra lines have been used. Zebra lines enabled identification of the smoothness criterion breach between two surfaces in almost all of the examples reviewed: object analysis after edge rounding in Inspire Studio, ANSYS SpaceClaim and Rhinoceros 6 identified sharp bends in the areas of contact between the rounded surfaces and the original surfaces (Figs. 6, 7, 8).

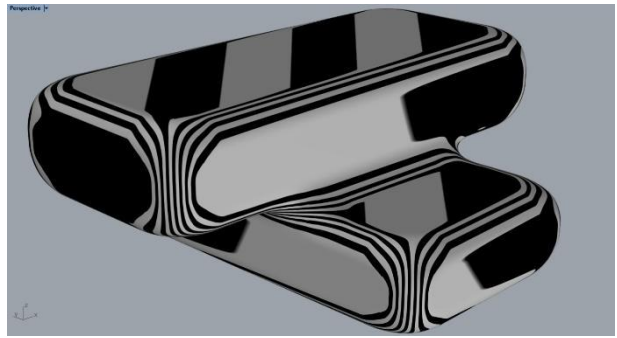

**Fig. 7.** Analysis of the surface of the designed object using zebra lines in Autodesk Inventor Professional 2020.

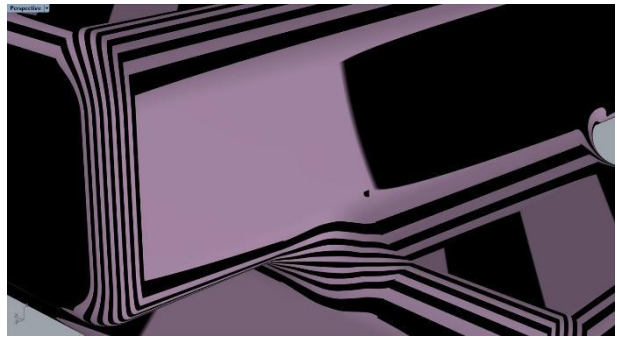

**Fig. 8.** Analysis of the surface of the designed object using zebra lines in ANSYS SpaceClaim.

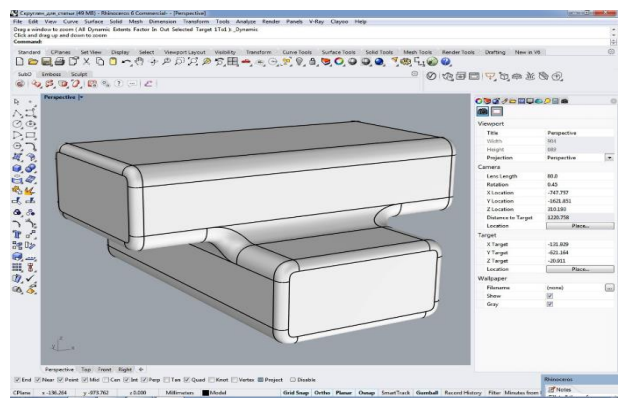

**Fig. 9.** Manually creating and editing edge rounding in Rhinoceros 6.

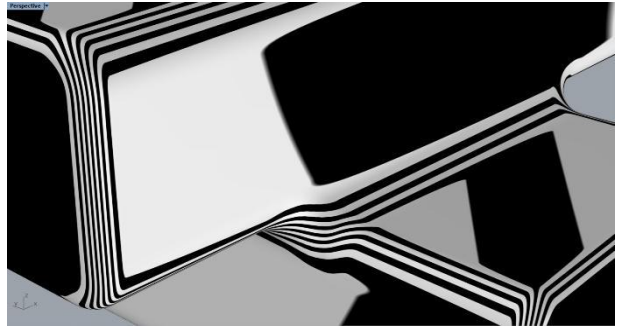

**Fig. 10.** Analysis of the surface of the designed object using zebra lines in Rhinoceros 6.

Thus, the considered software products could not cope with the creation of the rounded edges in the automatic mode, smoothly transferring one surface into another. Autodesk Inventor Professional 2020 created surfaces that met the criteria for smoothness, but at the same time created many unnecessary surfaces that complicated the topology of the object. A high-quality result, in which the zebra lines did not reveal excessively high curvature in the locations of mating surfaces, became possible with the manual edge rounding process (Fig. 10). Rhinoceros 6 has the best toolkit for this purpose.

The above analysis emphasizes the urgent need to improve the software kernel in order to automate many routine operations related to the modelling of industrial design products, one of which has been examined above.

## 6 Conclusion

This manuscript suggests a multi-criteria approach to assessing the quality of the shapes of functional curves that form surfaces, the quality of which substantially determines the functional characteristics of designed objects. The aesthetic functional curves are proposed to include the aesthetic curves that form the basis for shaping industrial design products and determining their consumer properties.

Based on the analysis of the motion of a point particle along a curved path, requirements for the quality of functional curves for the unstressed smooth motion of a point particle have been developed. A general list of quality requirements for the functional curves is defined (high order of smoothness, minimum number of extrema of the curvature, minimization of the maximum value of the curvature, minimization of the variation rate of the curvature, minimization of the potential energy of the curve). Additional requirements for aesthetic functional curves are defined from the standpoint of the laws of technical aesthetics. The curves satisfying the requirements for the functional curves are defined as class F curves.

To compare the quality of various CAD systems, an analysis of the rounded surfaces of the 3D model obtained in different computer-aided design systems according to the smoothness criterion $G^2$ has been carried out. The purpose of a visual demonstration by the method of comparative analysis of the low quality of fillets in various CAD systems was to show the imperfection of the mathematical apparatus of the geometric kernel in the software that was used.

## 7 Acknowledgements

We would like to thank the reviewers for their thoughtful comments and efforts towards improving our manuscript.